\documentclass[%
reprint,
superscriptaddress,
 amsmath,amssymb,
prl,
]{revtex4-2}

\usepackage{amssymb}
\usepackage{amsmath}
\usepackage{bbm}
\usepackage{bm}
\usepackage{epsfig}
\usepackage{color}
\usepackage{multirow}
\usepackage{graphicx}
\usepackage{booktabs}
\usepackage{dsfont}
\usepackage{tikz}
\usepackage{appendix}
\usepackage[colorlinks,
            linkcolor=blue,
            anchorcolor=green,
            citecolor=blue
            ]{hyperref}
\begin{document}
\title{Unconventional hybrid-order topological insulators} 
\author{Wei Jia}
\email[]{jiaw@lzu.edu.cn} 
\affiliation{Lanzhou Center for Theoretical Physics, Key Laboratory of Theoretical Physics of Gansu Province, Key Laboratory of Quantum Theory and Applications of MoE, Gansu Provincial Research Center for Basic Disciplines of Quantum Physics, Lanzhou University, Lanzhou 730000, China}
\author{Yuping Tian}
\affiliation{College of Sciences, Northeastern University, Shenyang 110819, China}
\author{Huanhuan Yang}
\email[]{yang.huanhuan.b6@tohoku.ac.jp}
\affiliation{Advanced Institute for Material Research (WPI-AIMR), Tohoku University, 2-1-1 Katahira, Sendai, 980-8577, Japan}
\author{Xiangru Kong}
\email[]{kongxiangru@mail.neu.edu.cn}
\affiliation{College of Sciences, Northeastern University, Shenyang 110819, China}
\author{Zhi-Hao Huang} 
\affiliation{Department of Physics, Fudan University, Shanghai 200433, China}
\author{Wei-Jiang Gong}
\affiliation{College of Sciences, Northeastern University, Shenyang 110819, China}
\author{Jun-Hong An}
\affiliation{Lanzhou Center for Theoretical Physics, Key Laboratory of Theoretical Physics of Gansu Province, Key Laboratory of Quantum Theory and Applications of MoE, Gansu Provincial Research Center for Basic Disciplines of Quantum Physics, Lanzhou University, Lanzhou 730000, China}

\begin{abstract}
Exploring novel topological matters with exotic quantum states has always been a core issue in the field of condensed matter physics, which can update the understanding of topological phases and broaden the classification of topological materials. Here, we report a class of unconventional hybrid-order topological insulators (HyOTIs), which simultaneously host various different higher-order topological states in a single band gap. Such topological states exhibit a unique bulk-boundary correspondence that is different from the well-known first-order topological states, higher-order topological states, and the coexistence of both. Particularly, we develop a generic surface theory to precisely capture them and discover a three-dimensional unconventional HyOTI protected by inversion symmetry, which renders both helical and corner topological states and exhibits {\color{black} an unprecedented} bulk-edge-corner correspondence. By adjusting the parameters of the system, we also observe the nontrivial phase transitions between the inversion-symmetric HyOTI and other conventional phases. We further propose a circuit-based experimental scheme to detect these interesting results. Remarkably, we demonstrate that a modified tight-binding model of bismuth can support the unconventional HyOTI, suggesting a possible route for its material realization. This work shall significantly advance the research of hybrid topological states in both theory and experiment.
\end{abstract}

\maketitle
{\it\color{blue}Introduction.}---Topological materials have always attracted great attention owing to their profound implications in fundamental physics and their promising potential for applications in quantum technologies~\cite{klitzing1980new,thouless1982quantized,wen1991non,kitaev2003fault,hasan2010colloquium}. 
First-order and higher-order topological insulators~\cite{qi2011topological,benalcazar2017quantized,liu2017novel,song2017d,langbehn2017reflection,schindler2018higher,sheng2019two,serra2018observation,peterson2018quantized}, as the prominent examples of topological materials, have demonstrated that a $d$-dimensional ($d$D) system can host $n$th-order topological states in its $(d-n)$D boundaries~\cite{bernevig2006quantum,zhang2013surface,ezawa2018topological,queiroz2019splitting,trifunovic2019higher,jia2021dynamically,zhang2021universal,yu2020high,LI20211502,PhysRevB.105.L041105,PhysRevB.106.245105,PhysRevB.105.L201301,benalcazar2022chiral,PhysRevB.110.L201117,zhang2025multiple}. However, this understanding that a single system only emerges one type of topological order has been broken by the discovery of conventional hybrid-order topological insulators (HyOTIs)~\cite{PhysRevB.99.125149,kooi2020hybrid,zhang2020symmetry}. These topological insulators show that first-order and second-order topological states coexist in the same or different band gaps, which have been experimentally observed in a $2$D phononic crystal~\cite{PhysRevLett.126.156801} and a $3$D realistic solid arsenic~\cite{hossain2024hybrid}. A common view holds that the conventional HyOTIs only allow the coexistence of both first-order and higher-order topological states~\cite{lai2024real,yang2024hybrid,l1n5-1jsm}.

It is important that the HyOTIs, based on their fundamental nature, may exhibit the coexistence of multiple higher-order topological states, namely unconventional HyOTIs. Even though a great deal of effort has been devoted in seeking  novel hybrid topological states~\cite{PhysRevB.101.121301,huang2024hybrid,PhysRevB.111.195107,tang2025coexisting}, such nontrivial states have never been found. A key challenge lies in how to effectively identify and exactly characterize them. This requires the development of a generic topological characterization theory for the unconventional hybrid topological states. On the other hand, the unconventional HyOTIs extend beyond the first-order, higher-order, and conventional hybrid-order topological states, which shall bring a novel bulk-boundary correspondence that is different from the conventional one, drive the exotic response to external field, and extend the classification of topological materials. Hence, revealing novel physical phenomena in the unconventional HyOTIs shall greatly update our recent understanding of the existing topological phases.

In this Letter, we report a class of unconventional HyOTIs, which can simultaneously host various different higher-order topological states in a $d$D gapped system. We further employ the effective mass fields on $(d-1)$D real-space surfaces to construct a set of topological indices, whose nonzero values provide the precise determination of their orders. By applying this theory, we discover a $3$D inversion-symmetric unconventional HyOTI, which hosts both helical and corner topological states in a single band gap. Such HyOTI is characterized by a $\mathbb{Z}_2$ topological invariant, manifesting as a bulk-edge-corner correspondence. 
We further propose a circuit-based scheme to detect these special properties. Finally, we predict that a modified tight-binding model of bismuth can support the inversion-symmetric unconventional HyOTI, providing an insight to seek its material candidate. These results are expected to advance both theoretical and experimental research on hybrid topological quantum states.

{\it\color{blue}Generic surface theory.}---Our starting point is a $d$D topological insulator hosting $s$th-order topological states on $(d-s)$D boundaries. Here $s$ takes two or more elements in the set $S=\{2,3,\cdots,n\}$ with $3\leqslant n\leqslant d$, implying that the different higher-order topological states can coexist in such system. The corresponding effective Hamiltonian with the boundaries reads
\begin{equation}\label{Hamiltonian_1}
\mathcal{H}(\mathbf{k},\mathbf{r})=\sum^{d}_{i=1} k_i\gamma_i+m_{1,\mathbf{r}}\Gamma_{1}+\sum^{n}_{j=2} m_{j,\mathbf{r}}\Gamma_{j}.
\end{equation} 
Here $\boldsymbol{\gamma}$ and $\boldsymbol{\Gamma}$ are the pseudospin operators and obey the anticommutation relations~\cite{morimoto2013topological,chiu2016classification}, which implies that $\mathcal{H}(\mathbf{k},\mathbf{r})$ is a Dirac-type Hamiltonian. It is observed that $\mathcal{H}(\mathbf{k},\mathbf{r})$ consists of three parts, of which the first part is $\mathbf{k}$-dependent linear dispersions to characterize the properties of Dirac cones. The remaining two parts are mass terms, allowing the position $\mathbf{r}$ dependence and are utilized to open energy gap of the Dirac cones, where $m_{1,\mathbf{r}}$ and $m_{j,\mathbf{r}}$ can be regarded as the effective bulk mass and boundary masses, respectively. We further assume $m_{1,\mathbf{r}}>0$ ($m_{1,\mathbf{r}}<0$) inside (outside) the topological insulator, and thus there are $m_{1,\mathbf{r}}=0$ and $m_{j,\mathbf{r}}\neq 0$ on the $(d-1)$D boundaries, inducing a nonzero edge energy gap. 
This means that the different higher-order topological states can coexist in $(d-s)$D boundaries with $s\geqslant 2$, manifesting as the unconventional HyOTIs. 

To exactly capture these hybrid topological states, we next study the $(d-1)$D boundary physics of the above system. By projecting the above effective Hamiltonian to the boundaries as $\tilde{\mathcal{H}}(\mathbf{k},\mathbf{r})=\mathbf{P}(\mathbf{r})\mathcal{H}(\mathbf{k},\mathbf{r})\mathbf{P}(\mathbf{r})$~\cite{benalcazar2017electric,PhysRevX.7.041069,PhysRevB.100.195135,hwang2019fragile,yan2019higher,PhysRevLett.122.187001,trifunovic2021higher,PRXQuantum.3.040312,jia2023unified,huang2024surface}, where $\mathbf{P}(\mathbf{r})=\left[1-\mathtt{i}(\mathbf{n}_{\mathbf{r}}\cdot\boldsymbol{\gamma})\Gamma_1\right]/2$ denotes the projection operator to the boundaries and $\mathbf{n}_\mathbf{r}$ is the surface normal vector at the position $\mathbf{r}$, we obtain an effective $(d-1)$D boundary Hamiltonian
\begin{equation}\label{Hamiltonian_surface}
\tilde{\mathcal{H}}(\mathbf{k},\mathbf{r})=\mathbf{k}_\text{b}\cdot\tilde{\boldsymbol{\gamma}}+\sum^{n}_{j=2} m_{j,\mathbf{r}}\tilde{\Gamma}_{j},
\end{equation}
where $\mathbf{k}_\text{b}=\mathbf{k}-(\mathbf{k}\cdot \mathbf{n}_\mathbf{r})\mathbf{n}_\mathbf{r}$ is the boundary momentum. Here $\tilde{\boldsymbol{\gamma}}$ and $\tilde{\boldsymbol{\Gamma}}$ are the boundary-projected Gamma matrices, which still obey the anticommutation relations. {\color{black}It means that $\tilde{\mathcal{H}}(\mathbf{k},\mathbf{r})$ is a Dirac-type Hamiltonian.} We observe that there are $n-1$ effective boundary masses $m_{j,\mathbf{r}}$, where $j=2,3,\cdots,n$. When the system presents $(d-s)$D topological states, these effective boundary masses shall host $q$ mass fields for each order $s$, i.e., $\mathbf{m}^{(s,q)}_{\mathbf{r}}=m_{j',\mathbf{r}}$ should induce mass defects with the dimension $d_\text{defect}=d-s$. Here $j'$ takes $s-1$ elements in the set $S$, and thus we have $q=1,2,\cdots,\mathbbm{C}^{s-1}_{n-1}$ with $\mathbbm{C}$ being combination. {\color{black}Note that these mass fields are also written as $\mathbf{m}^{(s,q)}_{\mathbf{r}}=\left(m^{(s,q)}_{1,\mathbf{r}},m^{(s,q)}_{2,\mathbf{r}},\cdots,m^{(s,q)}_{s-1,\mathbf{r}}\right)$}. Since $|\mathbf{m}^{(s,q)}_\mathbf{r}|$ is nonzero everywhere, we can define $q$ unit mass fields as $\mathbf{M}^{(s,q)}_{\mathbf{r}}={\mathbf{m}^{(s,q)}_\mathbf{r}}/{|\mathbf{m}^{(s,q)}_\mathbf{r}|}$. {\color{black}It is clear that $\mathbf{M}^{(s,q)}_{\mathbf{r}}=\left(M^{(s,q)}_{1,\mathbf{r}},M^{(s,q)}_{2,\mathbf{r}},\cdots,M^{(s,q)}_{s-1,\mathbf{r}}\right)$ is homotopic to a $(s-2)$D sphere $\mathsf{S}^{s-2}$.} By further defining $\mathbf{r}$ as a real-space parameter on $\mathsf{S}^{d_r}$ to surround the mass defects with codimension $d_c=(d-1)-d_\text{defect}$, the homotopy group $
\pi_{s-2}(\mathsf{S}^{d_r})=\mathbbm{Z}$ characterizes these mass defects~\cite{teo2010topological,shiozaki2014topology,geier2018second}. Here $d_r=d_c-1$ is associated with the codimension of mass defects in $(d-1)$D boundaries. 

\begin{figure}[t]
\centering
\includegraphics[width=1.0\columnwidth]{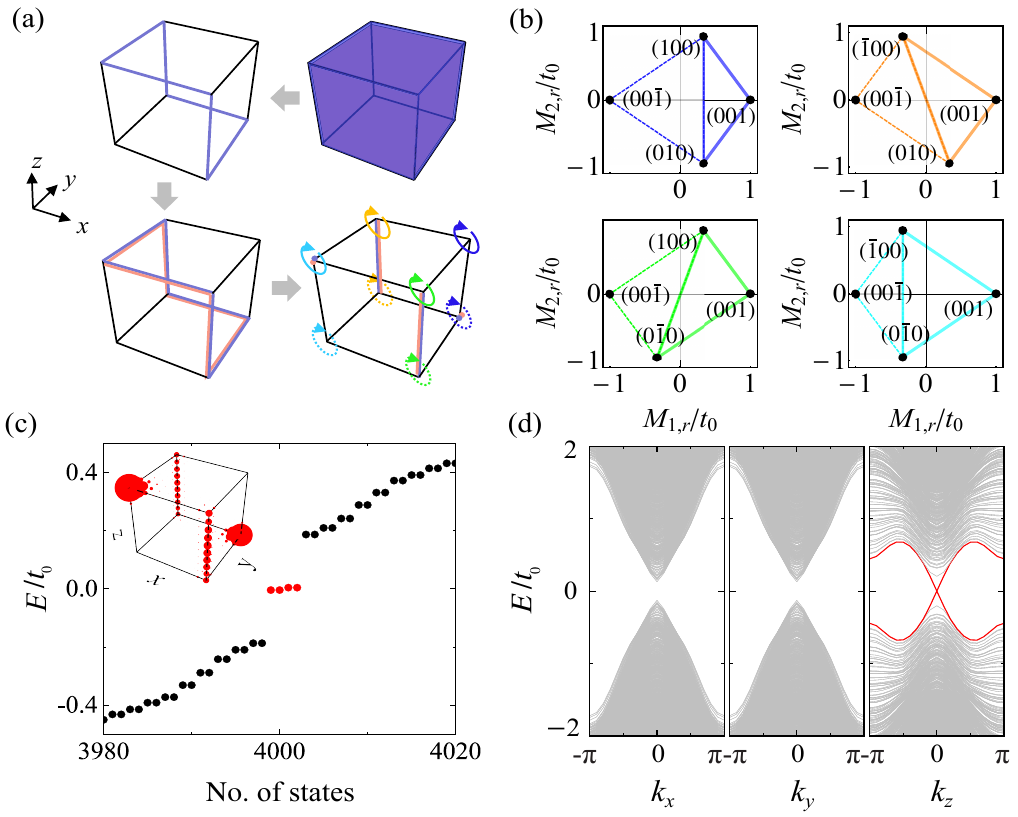}\caption{(a) The generation mechanism of a 3D inversion-symmetric unconventional HyOTI in a cubic crystal. (b) The distribution of unit mass fields enclosing 8 corners of (a). The bottom right and  top left corners host the mass defects, giving the nonzero $w^{(1)}_1$ and driving the corner states. The front and back edges host the mass defects, giving the nonzero $C^{(1),(2)}_0$ and driving the helical states. (c)-(d) The OBC energy spectrum gives four zero-energy states located in corners and hinges. The lattice size is $10\times 10\times 10$. The other parameters are $m_0=t_0$ and $B_0=0.35t_0$.}
\label{Fig:1}
\end{figure}

{\color{black}Under the above framework, we define a set of $\mathbb{Z}$ topological invariants to describe these mass defects, which is a $(s-2)$D winding number $w^{(q)}_{s-2}$ for the odd $s$ or the $\frac{s-2}{2}$th Chern number $C^{(q)}_{\frac{s-2}{2}}$ for the even $s$~\cite{chiu2016classification}. These nonzero invariants provide an elegant way to determine the orders $s$ of the unconventional HyOTI. One can further obtain
\begin{equation}\label{main}
\begin{split}
&C^{(q)}_0=\frac{1}{2}\left[\text{sgn}\left(M^{(2,q)}_{1,\mathbf{r}_\text{L}}\right)-\text{sgn}\left(M^{(2,q)}_{1,\mathbf{r}_\text{R}}\right)\right],\\
&w^{(q)}_1=\int \text{d}\theta\frac{1}{2\pi}\partial_\theta\tan{^{-1}}\left(\frac{M^{(3,q)}_{1,\theta}}{M^{(3,q)}_{2,\theta}}\right),\\
&C^{(q)}_1=\int\int \text{d}\theta\text{d}\phi\frac{1}{4\pi}\mathbf{M}^{(4,q)}_{\theta,\phi}\cdot\partial_\theta\mathbf{M}^{(4,q)}_{\theta,\phi}\times\partial_\phi\mathbf{M}^{(4,q)}_{\theta,\phi}, 
\end{split}
\end{equation}
and similar expressions in higher dimensions. Here $\mathbf{r}_\text{L(R)}$, $\theta$, and $\phi$ denote the real-space parameter paths surrounding the mass defects in $(d-s)$D interfaces. The details and their applications have been provided in the Supplementary Materials~\cite{Supplementary_Materials}. Besides, this surface theory is applicable for the non-Dirac-type topological system, where certain Gamma matrices in the Hamiltonian~\eqref{Hamiltonian_1} have commutation relations, as soon as its $(d-1)$D effective boundary Hamiltonian is Dirac-type.}  

{\it\color{blue} 3D inversion-symmetric HyOTI.}---Inspired by the surface theory, we discover a 3D inversion-symmetric unconventional HyOTI in a cubic crystal, with the coexistence of helical and corner states. The corresponding momentum-space Hamiltonian reads
\begin{equation}\label{Hamiltonian_HOHTP}
\begin{split}
\mathcal{H}(\mathbf{k})=\sum^3_{i=1}h_i\gamma_i+h_4\Gamma_1+\sum^3_{i=1}B_0\Gamma_{i+1}+h_5\Gamma_5,
\end{split}
\end{equation}
where $h$-components are given by $h_i=t_0\sin k_i$, $h_4=[m_0-t_0(3-\sum^3_{i=1}\cos k_i)]$, and $h_5=t_0(\cos k_1-\cos k_2)$. Here $t_0$ and $B_0$ are $\mathbf{k}$-independent constants and we take $k_{1,2,3}=k_{x,y,z}$. The Gamma matrices are $\gamma_1=\rho_3\tau_1\sigma_1$, $\gamma_2=\rho_3\tau_1\sigma_2$, $\gamma_3=\rho_3\tau_1\sigma_3$, $\Gamma_1=\rho_3\tau_3$, $\Gamma_2=\rho_3\sigma_1$, $\Gamma_3=\rho_3\sigma_2$, $\Gamma_4=\rho_3\sigma_3$, and $\Gamma_5=\rho_2$, where the Pauli matrices $\sigma_i$, $\tau_i$, and $\rho_i$ denote the freedom degree of spin, orbit, and sublattice, respectively. Note that $\mathcal{H}(\mathbf{k})$ only presents the inversion symmetry with ${\mathcal{I}}=\tau_3$ and is a non-Dirac type Hamiltonian due to $[\Gamma_{i+1},\gamma_i]=0$ and $[\Gamma_{2,3,4},\Gamma_{1}]=0$ with $i=1,2,3$. 

Physically, the $3$D inversion-symmetric HyOTI can be understood through four processes, as shown in Fig.~\ref{Fig:1}(a). Firstly, we consider a $3$D traditional $\mathbb{Z}_2$ topological insulator, described by $\mathcal{H}_0(\mathbf{k})=h_4\tau_3+\sum^3_{i=1}h_i\tau_1\sigma_i$~\cite{bernevig2006quantum}. In addition to ${\mathcal{I}}=\tau_3$, the system also presents time-reversal symmetry $\mathcal{T}=\mathtt{i}\sigma_2\mathcal{K}$ and four-fold rotation symmetry $C_{4}=\text{e}^{-\mathtt{i}\pi\sigma_3/4}$, which protect the existence of $2$D surface states. Secondly, we add a magnetic field $\mathbf{B}$ in the $\mathbb{Z}_2$ topological insulator, which induces the system to arrive $\mathcal{H}_0(\mathbf{k})+\mathcal{H}_1(\mathbf{k})$, with $\mathcal{H}_1(\mathbf{k})=\sum^3_{i=1} B_i\sigma_i$ and $B_i=B_0$. The nonzero $\mathbf{B}$ breaks $\mathcal{T}$ and $C_{4}$, thereby opening the energy gap of all surface states. Since ${\mathcal{I}}$ is still preserved and the surfaces perpendicular to $\mathbf{r}$ and $-\mathbf{r}$ have opposite magnetic flux, the chiral hinge states are emerged in the edges intersected by these surfaces~\cite{khalaf2018higher}. Thirdly, we perform the time-reversal copy for the above system with chiral hinge states. Then, it is described by $\rho_3[\mathcal{H}_0(\mathbf{k})+\mathcal{H}_1(\mathbf{k})]$, which drives these chiral hinge states to helical states. Finally, we add $\mathcal{H}_2(\mathbf{k})=h_5\rho_2$ to open the energy gap of helical states on $(001)$ and $(00\bar{1})$ surfaces, which induces the corner states. As $\mathcal{H}_2(\mathbf{k})$ is independent to $k_z$ and preserves the helical states along $z$ direction, we obtain the unconventional HyOTI with helical and corner states, as shown in Fig.~\ref{Fig:1}(a).

The surface theory further illustrates that the Hamiltonian~\eqref{Hamiltonian_HOHTP} can exhibit the HyOTI. By using the operator $\mathbf{P}(\mathbf{r})=\left[1-\mathtt{i}(\mathbf{n}_{\mathbf{r}}\cdot\boldsymbol{\gamma})\Gamma_1\right]/2$ to project $\mathcal{H}({\mathbf{k}})$ into six 2D surfaces of a cubic crystal, the effective surface Hamiltonians are easily obtained~\cite{Supplementary_Materials}. We examine their unit mass fields $\mathbf{M}_{\mathbf{r}}$ surrounding $8$ corners and $16$ edges [see Figs.~\ref{Fig:1}(a) and \ref{Fig:1}(b)]. {\color{black}For the corner intersected by $(001)$, $(0\bar{1}0)$, and $(\bar{1}00)$ surfaces, these unit mass fields figure out a path [see lightblue solid line in Fig.~\ref{Fig:1}(b)], which enclose the zero point and contribute $w^{(1)}_1=1$ in the parameter space of enclosing this corner~\cite{Supplementary_Materials}}. Then, we identify that there are $0$D topological states in this corner. {\color{black}For the edge intersected by $(100)$ and $(0\bar{1}0)$ surfaces, these unit mass fields figure out two paths, which cross the zero point [see green solid line in Fig.~\ref{Fig:1}(b)] and contribute $C^{(1)}_0=C^{(2)}_0=1$~\cite{Supplementary_Materials}}. Thus, there are $1$D topological states in this edge. Similarly, the remaining edges and corners are also identified, showing that this system simultaneously host second-order and third-order topological states~\cite{Supplementary_Materials}. Furthermore, we numerically obtain the energy spectrum of $\mathcal{H}(\mathbf{k})$ under open boundary conditions (OBCs), as shown in Figs.~\ref{Fig:1}(c) and~\ref{Fig:1}(d). It is seen that the helical states are located at two edges along $z$ direction and the corner states are located at one corner of the other two edges, which completely matches with the previous results of surface theory.          

\begin{figure}[!t]
\centering
\includegraphics[width=1.0\columnwidth]{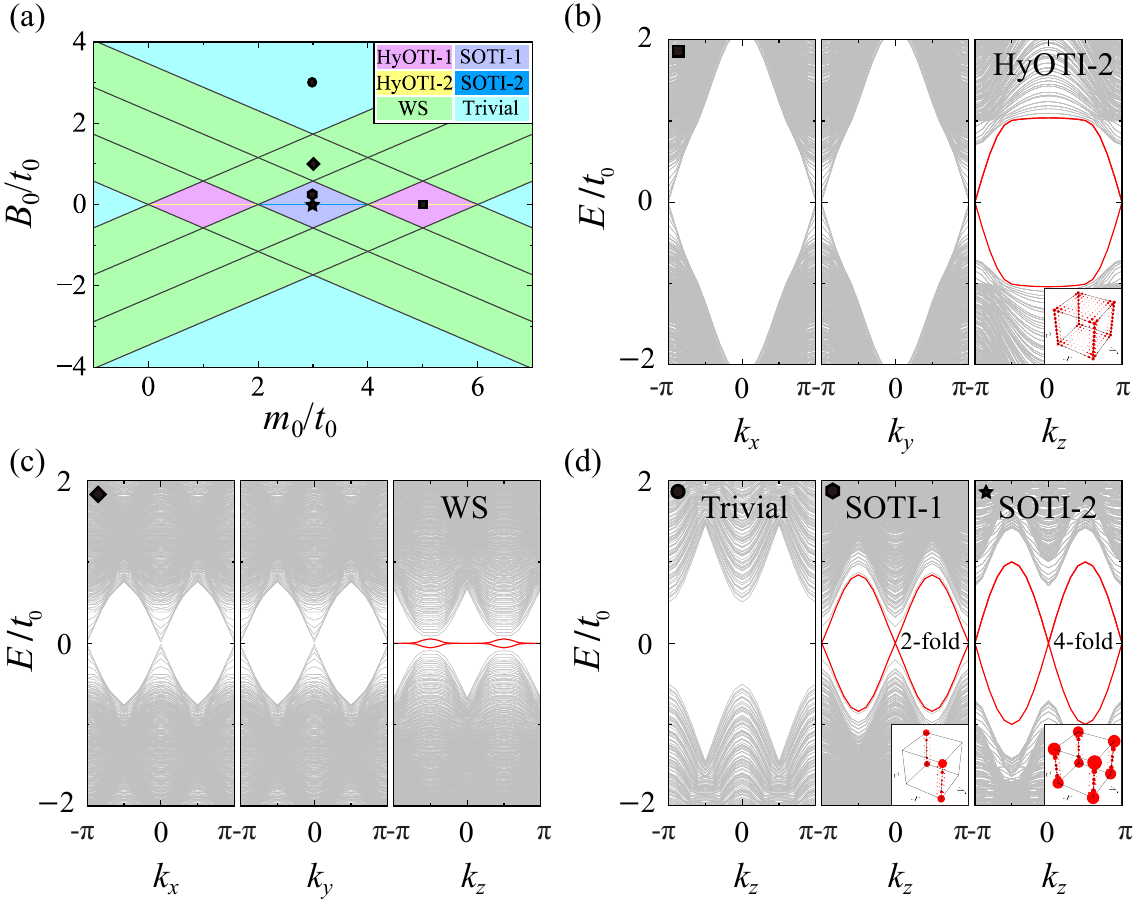}
\caption{Topological phase diagram of the model~\eqref{Hamiltonian_HOHTP}, including HyOTI-1, HyOTI-2, SOTI-1, SOTI-2, WS, and trivial phases. The OBC energy spectra are shown in (b) HyOTI-2, (c) WS, and (d) trivial phase, SOTI-1, and SOTI-2. The insets give the distribution of zero-energy states. Here the parameters are $m_0=5t_0$ and $B_0=0$ for (b), $m_0=3t_0$ and $B_0=t_0$ for (c), $m_0=3t_0$ and $B_0=3t_0$, $0.35t_0$, and $0$ for (d).}
\label{Fig:2}
\end{figure} 

{\it\color{blue}Topological phase diagram.}---The band structure of this $3$D inversion-symmetric HyOTI can be obtained by diagonalizing the Hamiltonian $\mathcal{H}(\mathbf{k})$. It allows that the phase boundaries of system are determined by closing bulk energy gap at four high-symmetry points $\mathbf{D}_1=(0,0,0)$, $\mathbf{D}_2=(0,0,-\pi)$, $\mathbf{D}_3=(-\pi,-\pi,0)$, and $\mathbf{D}_4=(-\pi,-\pi,-\pi)$. Namely, the phase transitions occur at $\sqrt{3}B_0=\pm m_0$, $\pm(m_0-2t_0)$, $\pm(m_0-4t_0)$, and $\pm(m_0-6t_0)$. Moreover, the phase transitions can emerge from $B_0=0$ to $B_0\neq 0$ due to the change of symmetries, although the bulk energy gap is kept. With this, we obtain the topological phase diagram in Fig.~\ref{Fig:2}(a). It is observed that there are six different phases: (i) unconventional HyOTI with two-fold helical and corner states (HyOTI-1); (ii) conventional HyOTI with four-fold helical states and $xy$-surface states (HyOTI-2); (iii) Weyl semimetal (WS); (iv) trivial phase; (v) SOTI with two-fold helical states (SOTI-1); (vi) SOTI with four-fold helical states (SOTI-2). By adjusting the strength of magnetic field, these phases render rich transitions. Particularly, there are two exotic types have not been reported so far, where one occurs between WS and HyOTI-1 by closing bulk energy gap, and the other occurs between HyOTI-1 and HyOTI-2 by changing the symmetries.    

\begin{figure}[!t]
\centering
\includegraphics[width=1.0\columnwidth]{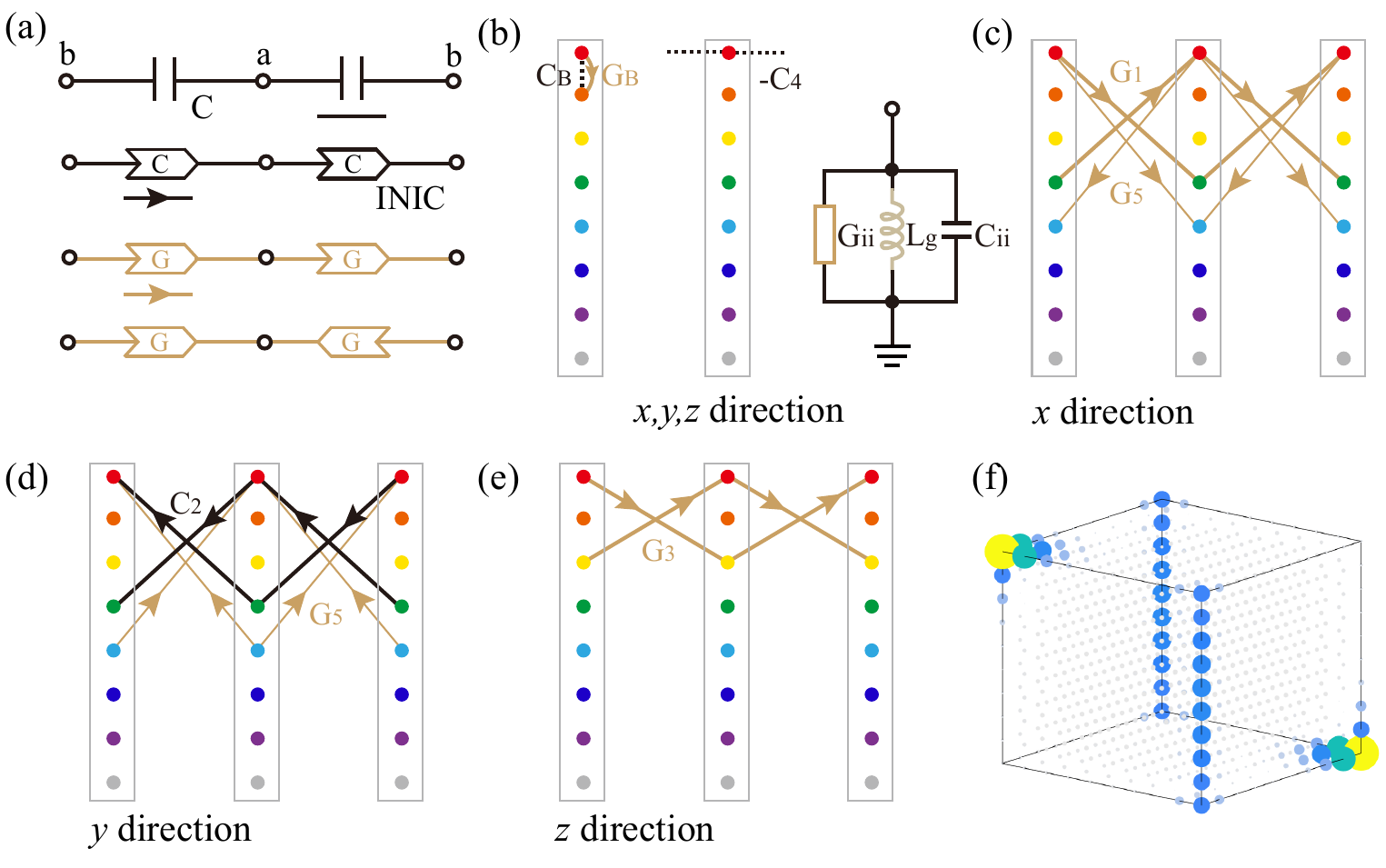}
\caption{(a) General circuit models for realizing distinct interactions. Here, INIC unit acts as a positive (negative) capacitor/resistor from right to left (left to right). (b) Circuit implementation of the magnetic field terms (left), hopping term $h_4$ (middle), and on-site potentials (right). (c)-(e) Other hopping terms along the $x$, $y$, and $z$ directions for first node. (f) Numerical impedance of a finite-size circuit model.}
\label{Fig:3}
\end{figure}

Besides, the above topological phase diagram implies that HyOTI-1 is induced by HyOTI-2, when changing the magnetic field $B_0$ from zero to nonzero. It allows that the topological characterization of HyOTI-2 is extended to HyOTI-1. Given that the inversion symmetry is always presented, we define a $\mathbb{Z}_2$ topological index 
\begin{equation}
\nu=\frac{\nu_+-\nu_-}{2}~~\text{mod}~~2
\end{equation}
to characterize both HyOTI-1 and HyOTI-2. Here $\nu_{\pm}$ are defined by $\mathcal{H}_{\pm}(\mathbf{k})$ which acts on the inversion subspace with even ($+$) and odd ($-$) parity, respectively. Since the rotoinversion symmetry with $\bar{C}_{4}=C_{4}\mathcal{I}$ are emerged for $\mathcal{H}_{\pm}(\mathbf{k})$ at $B_0=0$, we have
\begin{equation}
\nu_\pm=\frac{1}{2\sqrt{2}}\sum_{\mathbf{K}}\sum_\alpha e^{\frac{\mathtt{i}\alpha \pi}{4}}n^{\alpha,\pm}_\mathbf{K}.
\end{equation} 
Here $\mathbf{K}$ runs over the points $\mathbf{D}_{1,2,3,4}$ and $n^{\alpha,\pm}_\mathbf{K}$ is the number of the occupied bands of $\mathcal{H}_{\pm}(\mathbf{k})$ with the eigenvalue $\text{e}^{\frac{\mathtt{i}\alpha \pi}{4}}$ of the operator $\bar{C}_{4}$~\cite{khalaf2018symmetry}. After some straightforward calculations, we have $\nu=1$ for both HyOTI-1 and HyOTI-2, while $\nu=0$ is for the other phases. These results reveal a unique bulk-edge-corner (bulk-surface-edge) correspondence in HyOTI-1 (HyOTI-2).

\begin{figure}[!b]
\centering
\includegraphics[width=1.0\columnwidth]{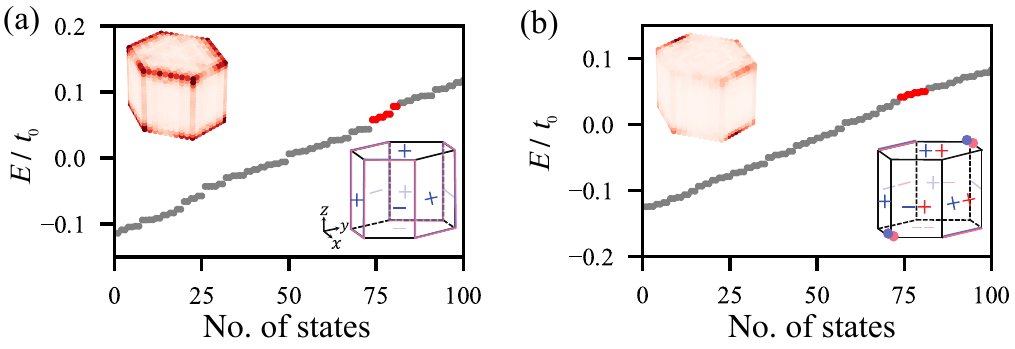}
\caption{The OBC energy spectrum of the tight-binding model (a) without and (b) with additional mass terms on a hexagonal structure. The 100 eigenstates near the Fermi level are plotted, where two insets illustrate the real-space contributions of the states with red color and the signs of effective surface masses, respectively. The system size is $20\times 20\times 20$.}
\label{Fig:4}
\end{figure}

{\it \color{blue} Circuit-based realization.}---Topological circuits constitute a versatile platform for simulating lattice Hamiltonians with nontrivial band topology~\cite{PhysRevLett.126.146802,wang2020circuit,yu20204d,PhysRevLett.132.046601,10.1063/5.0265293}. Here, we realize the inversion-symmetric HyOTI by mapping the Bloch Hamiltonian~\eqref{Hamiltonian_HOHTP} onto a frequency-dependent circuit Laplacian, i.e., $\mathcal{H}(\mathbf{k})\rightarrow-\mathtt{i}\mathcal{J}(\omega)$, following Refs.\cite{helbig2020generalized,yang2024circuit}. To reproduce the distinct hopping amplitudes in $\mathcal{H}(\mathbf{k})$, we employ the general circuit cell shown in Fig.~\ref{Fig:3}(a). For an infinitely extended chain its admittance elements are $j_{ab}=\mathtt{i}[-2\omega C\cos k]$, $j_{ab}=\mathtt{i}[2\mathtt{i}\omega C\sin k]$, $j_{ab}=\mathtt{i}[2G\sin k]$, and $j_{ab}=\mathtt{i}[2\mathtt{i}G\cos k]$, respectively, where $C$ ($G$) denotes the capacitance (conductance) and $k$ is the momentum along the chain. 

Using these building blocks we construct a 3D cubic network whose unit cell contains 8 circuit nodes connected by capacitors ($C$), inductors ($L$), resistors ($R=1/G$), and negative impedance converter with current inversion (INICs), as sketched in Figs.~\ref{Fig:3}(b)-(e). For clarity, only the couplings from the first (red) node are shown. The full $8\times8$ Laplacian matrix $\mathcal J(\omega)$, which exactly implements the Hamiltonian $\mathcal{H}(\mathbf{k})$, is given in the Supplemental Material~\cite{Supplementary_Materials}. For a finite sample with $N_x=N_y=N_z=10$ unit cells, we assemble the corresponding Laplacian and compute the impedance between each node and the ground numerically. The result, displayed in Fig.~\ref{Fig:3}(f), exhibits simultaneous boundary resonances characteristic of second-order and third-order topological states, thereby confirming our hybrid-order theoretical framework. By further adjusting the parameters of the circuit, we can realize all the phases in Fig.~\ref{Fig:3}(a) and observe these nontrivial phase transitions. This scheme provides a feasible experimental way to detect the nontrivial properties emerged in the inversion-symmetric unconventional HyOTI.

{\it \color{blue} Possible material candidates.}---We next show that the $3$D unconventional HyOTI can be observed in a hexagonal lattice hosting the tight-binding Hamiltonian
\begin{equation}
    \mathcal{H}_\text{TB}(\mathbf{k})=
    \begin{bmatrix}
      \mathcal{H}_\text{I}(\mathbf{k})+m_z & \delta \mathcal{M}(\mathbf{k})+m_x\\
    \delta \mathcal{M}(\mathbf{k})^{\dagger}+m_x&   \mathcal{H}_\text{II}(\mathbf{k})-m_z \\
    \end{bmatrix}.
\end{equation}
When the additional mass terms $m_{x,z}$ vanish, it is seen that $\mathcal{H}_\text{TB}(\mathbf{k})$ consists of two 3D topological insulators given by $\mathcal{H}_\text{I}(\mathbf{k})$ and $\mathcal{H}_\text{II}(\mathbf{k})$ and they couple together via the mass matrix $\mathcal{M}(\mathbf{k})$ with the coupling strength $\delta$~\cite{Supplementary_Materials}. Then, $\mathcal{H}_\text{TB}(\mathbf{k})$ is topologically equivalent to the realistic model of pure bismuth~\cite{PhysRevB.52.1566,NJP123015}, which renders a SOTI with helical states. The OBC energy spectrum is shown in Fig.~\ref{Fig:4}(a), where 8 states marked as red contribute at the edges. We emphasize that such helical states are determined by the mass domain walls induced by effective surface masses [see the inset of Fig.~\ref{Fig:4}(a)]. 

However, the nonzero $m_{x,z}$ can change the effective surface masses into the forms of the inset of Fig.~\ref{Fig:4}(b), resulting in the unconventional HyOTI~\cite{Supplementary_Materials}. The OBC energy spectrum in Fig.~\ref{Fig:4}(b) confirms that the system host both helical and corner states. Note that the forms of additional mass terms can be arbitrary, but the system can show such unconventional HyOTI as soon as the effective surface mass field meets the requirements of Eq.~\eqref{main}. These results shall provide a possible route to seek the realistic unconventional HyOTI in bismuth-based materials.

{\it \color{blue} Discussion and Conclusion.}---Actually, the unconventional HyOTIs may also emerge in non-Hermitian and Floquet systems, since the coexistence of first-order and second-order topological states has been observed in these systems~\cite{PhysRevB.109.134302,PhysRevLett.134.176601,PhysRevLett.123.016805,PhysRevB.107.235132,PhysRevB.110.235140}. Therefore, such hybrid topological states can emerge in the broader topological systems. Moreover, the recent experiments have demonstrated that pure bismuth is a SOTI with helical states~\cite{10.1038/s41567-018-0224-7,zhao2025revealing}. We have predicted that the tight-binding model of bismuth can support the unconventional HyOTI by adjusting the relative energy shift (i.e., $m_z$) between two topological insulator layers and changing their coupling constant (i.e., $m_x$). In the realistic materials, a possible route to achieving the unconventional HyOTIs involves doping bismuth with magnetic elements~\cite{hofmann2006surfaces,hegde2022review} or interfacing with magnetic substrates~\cite{PhysRevB.107.104428,zhou2023spin}. Both of two processes can tune the effective surface masses to satisfy the requirements of the surface theory. 

In summary, we have revealed a class of unconventional HyOTIs, which simultaneously host various different higher-order topological states in a single system and exhibits a unique bulk-boundary correspondence. We have developed a generic theory to exactly capture their nontrivial physics and proposed a circuit-based feasible scheme to detect the interesting phenomena in the unconventional HyOTIs. Besides, we have predicted that a modified tight-binding model of bismuth can support the unconventional HyOTI. This work is expected to advance both theoretical and experimental research on hybrid topological states.

{\it \color{blue} Acknowledgements.}---We thank Dr. Zhi-Xiong Li and Prof. Xiong-Jun Liu for the helpful discussions. W. Jia is supported by the National Natural Science Foundation of China (Grant No. 12404318), the Fundamental Research Funds for the Central Universities (Grant No. lzujbky-2024-jdzx06), the Natural Science Foundation of Gansu Province (No. 22JR5RA389), and the ‘111 Center’ under Grant No. B20063. X. Kong is supported by the Fundamental Research Funds for the Central Universities (No. N25LPY025).

\bibliography{references}

\pagebreak
\clearpage
\onecolumngrid
\flushbottom
\begin{center}
\textbf{\large Supplementary Material for ``Unconventional hybrid-order topological insulators"}
\end{center}
\setcounter{equation}{0}
\setcounter{figure}{0}
\setcounter{table}{0}
\makeatletter
\renewcommand{\theequation}{S\arabic{equation}}
\renewcommand{\thefigure}{S\arabic{figure}}
\renewcommand{\bibnumfmt}[1]{[S#1]}
\renewcommand{\citenumfont}[1]{S#1}

In this Supplementary Material, we provide the instruction of surface theory for a $4$D HyOTI in Sec.~I. We further provide the details of effective surface Hamiltonians of 3D inversion-symmetric HyOTI in Sec.~II. We show the more applications of the surface theory in Sec.~III. We also provide the details of circuit-based scheme for inversion-symmetric HyOTIs in Sec.~IV. We finally show the completely numerical results of the tight-binding model of bismuth with additional mass terms in Sec.~V. 

\begin{figure}[!t]
\centering
\includegraphics[width=0.5\columnwidth]{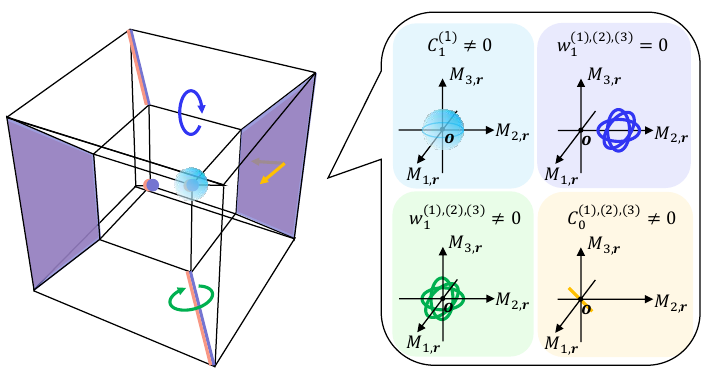}
\caption{{\color{black}Schematic diagram of a $4$D HyOTI with surface, hinge, and corner states, which is characterized by the nonzero $C^{(1),(2),(3)}_0$, $w^{(1),(2),(3)}_1$, and $C^{(1)}_1$, respectively.}}
\label{Fig:0}
\end{figure}

{\color{black}
\subsection{I. Instruction of surface theory for a 4D HyOTI}\label{Surface theory}

We consider a $4$D HyOTI with $d=n=4$ to illustrate the surface theory. This topological system is described by the bulk Hamiltonian
\begin{equation}\label{Hamiltonian_1}
\mathcal{H}(\mathbf{k},\mathbf{r})=\sum^{d}_{i=1} k_i\gamma_i+m_{1,\mathbf{r}}\Gamma_{1}+\sum^{4}_{j=2} m_{j,\mathbf{r}}\Gamma_{j}.
\end{equation}
It is clear that there are three effective boundary masses $m_{2,\mathbf{r}}$, $m_{3,\mathbf{r}}$, and $m_{4,\mathbf{r}}$ on the 3D surfaces. When taking $S=\{2,3,4\}$, this $4$D HyOTI can simultaneously host surface (i.e., $s=2$), hinge (i.e., $s=3$), and corner states (i.e., $s=4$), as shown in Fig.~\ref{Fig:0}. For $s=2$, the effective $1$D mass fields are given by $\mathbf{m}^{(2,1)}_{\mathbf{r}}=m^{(2,1)}_{1,\mathbf{r}}=m_{2,\mathbf{r}}$, $\mathbf{m}^{(2,2)}_{\mathbf{r}}=m^{(2,2)}_{1,\mathbf{r}}=m_{3,\mathbf{r}}$, and $\mathbf{m}^{(2,3)}_{\mathbf{r}}=m^{(2,3)}_{1,\mathbf{r}}=m_{4,\mathbf{r}}$. The corresponding $1$D unit mass fields read
\begin{equation}
\begin{split}
&\mathbf{M}^{(2,1)}_{\mathbf{r}}=M^{(2,1)}_{1,\mathbf{r}}=m_{2,\mathbf{r}}/|m_{2,\mathbf{r}}|\equiv M_{1,\mathbf{r}},\\
&\mathbf{M}^{(2,2)}_{\mathbf{r}}=M^{(2,2)}_{1,\mathbf{r}}=m_{3,\mathbf{r}}/|m_{3,\mathbf{r}}|\equiv M_{2,\mathbf{r}},\\
&\mathbf{M}^{(2,3)}_{\mathbf{r}}=M^{(2,3)}_{1,\mathbf{r}}=m_{4,\mathbf{r}}/|m_{4,\mathbf{r}}|\equiv M_{3,\mathbf{r}},
\end{split} 
\end{equation}
and can be used to define the zeroth Chern number
\begin{equation}
\begin{split}
C^{(q)}_0=\left[\text{sgn}\left(M^{(2,q)}_{1,\mathbf{r}_\text{L}}\right)-\text{sgn}\left(M^{(2,q)}_{1,\mathbf{r}_\text{R}}\right)\right]/2,
\end{split}
\end{equation}
where $\mathbf{r}_\text{L}$ and $\mathbf{r}_\text{R}$ are denoted in two $3$D surfaces (see two orange arrows in Fig.~\ref{Fig:0}), which can cross to product the $2$D interface. The existence of $2$D surface states are determined by the non-zero $C^{(1),(2),(3)}_0$ (see Fig.~\ref{Fig:0}). For $s=3$, these $1$D mass fields are written as $\mathbf{m}^{(3,1)}_{\mathbf{r}}=\left(m^{(3,1)}_{1,\mathbf{r}},m^{(3,1)}_{2,\mathbf{r}}\right)=(m_{2,\mathbf{r}},m_{3,\mathbf{r}})$, $\mathbf{m}^{(3,2)}_{\mathbf{r}}=\left(m^{(3,2)}_{1,\mathbf{r}},m^{(3,2)}_{2,\mathbf{r}}\right)=(m_{3,\mathbf{r}},m_{4,\mathbf{r}})$, and $\mathbf{m}^{(3,3)}_{\mathbf{r}}=\left(m^{(3,3)}_{1,\mathbf{r}},m^{(3,3)}_{2,\mathbf{r}}\right)=(m_{4,\mathbf{r}},m_{2,\mathbf{r}})$. The corresponding $2$D unit mass fields read
\begin{equation}
\begin{split}
& \mathbf{M}^{(3,1)}_{\mathbf{r}}=\left(M^{(3,1)}_{1,\mathbf{r}},M^{(3,1)}_{2,\mathbf{r}}\right)=\left(m_{2,\mathbf{r}}/\sqrt{m^2_{2,\mathbf{r}}+m^2_{3,\mathbf{r}}},m_{3,\mathbf{r}}/\sqrt{m^2_{2,\mathbf{r}}+m^2_{3,\mathbf{r}}}\right)\equiv (M_{1,\mathbf{r}},M_{2,\mathbf{r}}),\\
& \mathbf{M}^{(3,2)}_{\mathbf{r}}=\left(M^{(3,2)}_{1,\mathbf{r}},M^{(3,2)}_{2,\mathbf{r}}\right)=\left(m_{3,\mathbf{r}}/\sqrt{m^2_{3,\mathbf{r}}+m^2_{4,\mathbf{r}}},m_{4,\mathbf{r}}/\sqrt{m^2_{3,\mathbf{r}}+m^2_{4,\mathbf{r}}}\right)\equiv (M_{2,\mathbf{r}},M_{3,\mathbf{r}}),\\
&\mathbf{M}^{(3,3)}_{\mathbf{r}}=\left(M^{(3,3)}_{1,\mathbf{r}},M^{(3,3)}_{2,\mathbf{r}}\right)=\left(m_{4,\mathbf{r}}/\sqrt{m^2_{2,\mathbf{r}}+m^2_{4,\mathbf{r}}},m_{2,\mathbf{r}}/\sqrt{m^2_{2,\mathbf{r}}+m^2_{4,\mathbf{r}}}\right)\equiv (M_{3,\mathbf{r}},M_{1,\mathbf{r}})
\end{split}
\end{equation}
and can be used to define the $1$D winding number
\begin{equation}
\begin{split}
w^{(q)}_1=\int \text{d}\theta\frac{1}{2\pi}\partial_\theta\tan{^{-1}}\left({M^{(3,q)}_{1,\theta}}/{M^{(3,q)}_{2,\theta}}\right),
\end{split}
\end{equation}
where $\theta$ denotes a closed $1$D parameter path (see blue or green arrow in Fig.~\ref{Fig:0}) on the $3$D surfaces to surround the $1$D interface. The existence of $1$D edge states are determined by the non-zero $w^{(1),(2),(3)}_1$ (see Fig.~\ref{Fig:0}). For $s=4$, the $3$D mass fields are written as $\mathbf{m}^{(4,1)}_{\mathbf{r}}=\left(m^{(4,1)}_{1,\mathbf{r}},m^{(4,1)}_{2,\mathbf{r}},m^{(4,1)}_{3,\mathbf{r}}\right)=(m_{2,\mathbf{r}},m_{3,\mathbf{r}},m_{4,\mathbf{r}})$. The $3$D unit mass field reads 
\begin{equation}
\begin{split}
\mathbf{M}^{(4,1)}_{\mathbf{r}}&=\left(M^{(4,1)}_{1,\mathbf{r}},M^{(4,1)}_{2,\mathbf{r}},M^{(4,1)}_{3,\mathbf{r}}\right)\\
&=\left(m_{2,\mathbf{r}}/\sqrt{m^2_{2,\mathbf{r}}+m^2_{3,\mathbf{r}}+m^2_{4,\mathbf{r}}},m_{3,\mathbf{r}}/\sqrt{m^2_{2,\mathbf{r}}+m^2_{3,\mathbf{r}}+m^2_{4,\mathbf{r}}},m_{4,\mathbf{r}}/\sqrt{m^2_{2,\mathbf{r}}+m^2_{3,\mathbf{r}}+m^2_{4,\mathbf{r}}}\right)\\
&\equiv(M_{1,\mathbf{r}},M_{2,\mathbf{r}},M_{3,\mathbf{r}}),
\end{split}
\end{equation}
and defines the first Chern number
\begin{equation}
\begin{split}
C^{(1)}_1=\int\int \text{d}\theta\text{d}\phi\frac{1}{4\pi}\mathbf{M}^{(4,1)}_{\theta,\phi}\cdot\partial_\theta\mathbf{M}^{(4,1)}_{\theta,\phi}\times\partial_\phi\mathbf{M}^{(4,1)}_{\theta,\phi}, 
\end{split}
\end{equation}
where $\theta$ and $\phi$ denote a closed $2$D parameter surface (see lightblue sphere in Fig.~\ref{Fig:0})) on the $3$D surfaces to surround the $0$D interface. The existence of $0$D corner states are determined by non-zero $C^{(1)}_1$ (see Fig.~\ref{Fig:0}).}

\subsection{II. Effective surface Hamiltonians of inversion-symmetric HyOTI in a cubic geometry}\label{Effective surface Hamiltonian}
We consider a $3$D momentum-space Hamiltonian, which can drive the inversion-symmetric HyOTI with helical and corner states in a cubic crystal. Its general form is
\begin{equation}\label{s_model}
\mathcal{H}(\mathbf{k})=\left[m_0-t_0\left(3-\sum^3_{i=1}\cos k_i\right)\right]\rho_3\tau_3\sigma_0+t_0\sum^3_{i=1}\sin k_i\rho_3\tau_1\sigma_i+\sum^3_{i=1} B_i\rho_3\tau_0\sigma_i +t_0\left(\cos k_1-\cos k_2\right)\rho_2\tau_0\sigma_0.
\end{equation}
Here the Pauli matrices $\sigma_i$, $\tau_i$, and $\rho_i$ denote the freedom degree of spin, orbit, and sublattice, respectively. Note that $\rho_{0}$, $\tau_{0}$, and $\sigma_{0}$ are identity matrices. This Hamiltonian $\mathcal{H}(\mathbf{k})$ is same with Eq.~(5) of the main text when taking $B_{1,2,3}=B_0$. By taking $\mathbf{k}\rightarrow \mathbf{0}$, its low-energy effective Hamiltonian reads
\begin{equation}
\mathcal{H}_\text{eff}(\mathbf{k})=\left[m_0-t_0\left(\frac{1}{2}k_1^2+\frac{1}{2}k_2^2+\frac{1}{2}k_3^2\right)\right]\rho_3\tau_3\sigma_0+t_0\sum_{i}k_i\rho_3\tau_1\sigma_i+\sum_i B_i\rho_3\tau_0\sigma_i+t_0\left(\frac{1}{2}k_2^2-\frac{1}{2}k_1^2\right)\rho_2\tau_0\sigma_0.
\end{equation}
After using the projection operator 
\begin{equation}
\mathbf{P}_i=\frac{1}{2}(1-\mathtt{i}\rho_3\tau_3\sigma_0\cdot\rho_3\tau_1\sigma_i)
\end{equation}
in the eigenspace of $\mathtt{i}\rho_3\tau_3\sigma_0\cdot\rho_3\tau_1\sigma_i=-1$ and projecting $\mathcal{H}_\text{eff}(\mathbf{k})$ along $k_i$ with $i=1,2,3$, we obtain six effective surface Hamiltonians in a cubic geometry,
\begin{equation}
\begin{split}
&\tilde{\mathcal{H}}_{(001)}=\tilde{\mathcal{H}}_\text{I}(k_1,k_2)=t_0k_1\rho_3\sigma_3-t_0k_2\rho_3\sigma_1+B_3\rho_3\sigma_2,\\
&\tilde{\mathcal{H}}_{(010)}=\tilde{\mathcal{H}}_\text{II}(k_1,k_3)=t_0k_1\rho_3\sigma_3+t_0k_3\rho_3\sigma_1+B_2\rho_3\sigma_2-m_0\rho_2\sigma_0,\\
&\tilde{\mathcal{H}}_{(100)}=\tilde{\mathcal{H}}_\text{III}(k_2,k_3)=-t_0k_2\rho_3\sigma_3+t_0k_3\rho_3\sigma_1
+B_1\rho_3\sigma_2+{m_0}\rho_2\sigma_0,\\
&\tilde{\mathcal{H}}_{(00\bar{1})}=\tilde{\mathcal{H}}_\text{IV}(k_1,k_2)=-t_0k_1\rho_3\sigma_3+t_0k_2\rho_3\sigma_1-B_3\rho_3\sigma_2,\\
&\tilde{\mathcal{H}}_{(0\bar{1}0)}=\tilde{\mathcal{H}}_\text{V}(k_1,k_3)=-t_0k_1\rho_3\sigma_3-t_0k_3\rho_3\sigma_1-B_2\rho_3\sigma_2-m_0\rho_2\sigma_0,\\
&\tilde{\mathcal{H}}_{(\bar{1}00)}=\tilde{\mathcal{H}}_\text{VI}(k_2,k_3)=t_0k_2\rho_3\sigma_3-t_0k_3\rho_3\sigma_1-B_1\rho_3\sigma_2+m_0\rho_2\sigma_0.\\
\end{split}
\label{eff}
\end{equation}
Here the subscripts of $\tilde{\mathcal{H}}$ denote the Miller index. Note that the above surface Hamiltonians have been rotated without changing their topologies, so that the effective surface mass fields are easily obtained as follows: 
\begin{equation}
\begin{split}
&\mathbf{m}_\text{I}=(B_3,0),~~~\mathbf{m}_\text{II}=(B_2,-m_0),~~~\mathbf{m}_\text{III}=(B_1,m_0),\\&\mathbf{m}_\text{IV}=(-B_3,0),~\mathbf{m}_\text{V}=(-B_2,-m_0),~\mathbf{m}_\text{VI}=(-B_1,m_0).
\end{split}
\end{equation}
It is seen that each surface hosts two effective masses, as shown in Fig.~\ref{Fig:s1}(a). Due to the inversion symmetry, the surfaces I, II, and III have opposite magnetic flux with the surfaces IV, V, and VI. Namely, we have $B_i$ for I, II, and III surfaces but $-B_i$ for IV, V, and VI surfaces.

\subsection{III. More applications of the surface theory}

{\color{black} We next provides the detailed applications of surface theory for the Hamiltonians~\eqref{eff}, where the parameters $B_{1,2,3}=0.35t_0$ and $m_0=t_0$ are taken. To exactly capture the hybrid-order topological states, we need to consider all corners and edges of this cubic crystal. The corresponding topological indexes are then calculated. Firstly, we identify the unit mass fields surrounding all corners, given by $\mathbf{M}^{(3,1)}_{\mathbf{r}}=(M_{1,\mathbf{r}},M_{2,\mathbf{r}})$, as shown in Fig.~\ref{Fig:s1}(b). For example of the corner intersected by I, V, and VI surfaces, we obtain the unit mass fields as
\begin{equation}
\mathbf{M}^{(3,1)}_\text{I}=(1,0),~~~\mathbf{M}^{(3,1)}_\text{V}=\left(-\frac{B_2}{\sqrt{B_2^2+m_0^2}},-\frac{m_0}{\sqrt{B_2^2+m_0^2}}\right),~~~\mathbf{M}^{(3,1)}_\text{VI}=\left(-\frac{B_1}{\sqrt{B_1^2+m_0^2}},\frac{m_0}{\sqrt{B_1^2+m_0^2}}\right).
\end{equation}
These values figure out a path, which encloses the zero point and gives a nonzero winding number $w_\text{I,V,VI}=1$, as shown in Fig.~\ref{Fig:s1}(b) and \ref{Fig:s1}(c). It implies that this 2D unit mass field can figure out a nonzero winding number $w^{(1)}_1=w_\text{I,V,VI}=1$ along a 1D closed parameter path $\theta$ surrounding the corner, revealing that there are zero-energy states in this corner. For example of the corner intersected by I, II, and III surfaces, we obtain the unit mass fields as
\begin{equation}
\mathbf{M}^{(3,1)}_\text{I}=(1,0),~~~\mathbf{M}^{(3,1)}_\text{II}=\left(\frac{B_2}{\sqrt{B_2^2+m_0^2}},-\frac{m_0}{\sqrt{B_2^2+m_0^2}}\right),~~~\mathbf{M}^{(3,1)}_\text{III}=\left(\frac{B_1}{\sqrt{B_1^2+m_0^2}},\frac{m_0}{\sqrt{B_1^2+m_0^2}}\right).
\end{equation}
These values figure out a path, which can not enclose the zero point and gives a zero winding number $w_\text{I,II,III}=0$, as shown in Fig.~\ref{Fig:s1}(b) and \ref{Fig:s1}(c). It implies that this 2D unit mass field figures out a zero winding number $w^{(1)}_1=w_\text{I,II,II}=0$ along a 1D closed parameter path $\theta$ surrounding the corner, revealing that there is no zero-energy state in this corner. Similarly, we can identify $w^{(1)}_1$ for the remaining corners, the completely results are shown in Fig.~\ref{Fig:s1}(b). 

\begin{figure}[!t]
\centering
\includegraphics[width=0.98\columnwidth]{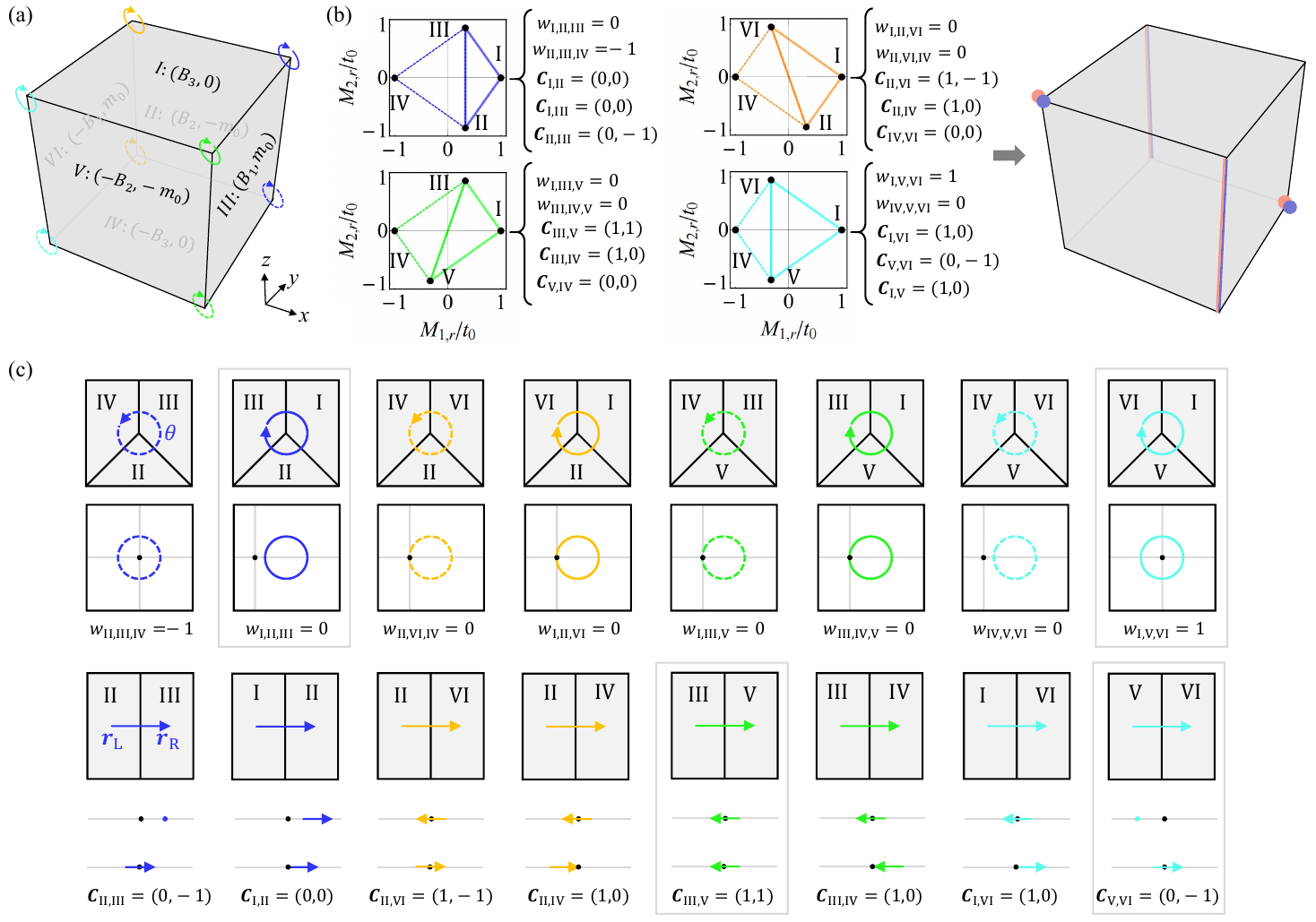}
\caption{(a) Distribution of effective masses on  $6$ surfaces of a cubic crystal. (b) A $3$D unconventional HyOTI with corner and helical states can be identified by the nonzero $w^{(1)}_1$ and $C^{(1),(2)}_0$, respectively. These topological invariants are defined by the unit surface mass fields $\mathbf{M}_{\mathbf{r}}$.
{\color{black}(c) Detailed calculations of these topological invariants for the different corners and edges.}}
\label{Fig:s1}
\end{figure}

Next, we calculate the topological indexes for all edges. The mass fields surrounding all $12$ edges are $\mathbf{M}^{(2,1)}_{\mathbf{r}}=M_{1,\mathbf{r}}$ and $\mathbf{M}^{(2,2)}_{\mathbf{r}}=M_{2,\mathbf{r}}$, as shown in Fig.~\ref{Fig:s1}(b). For example of the edge intersected III and V surfaces, we obtain the non-unit mass fields as 
\begin{equation}
\begin{split}
\mathbf{M}^{(2,1)}_\text{III}=\frac{B_1}{\sqrt{B_1^2+m_0^2}},~\mathbf{M}^{(2,1)}_\text{V}=-\frac{B_2}{\sqrt{B_2^2+m_0^2}},~\mathbf{M}^{(2,2)}_\text{III}=\frac{m_0}{\sqrt{B_1^2+m_0^2}},~\mathbf{M}^{(2,2)}_\text{V}=-\frac{m_0}{\sqrt{B_2^2+m_0^2}}.
\end{split}
\end{equation}
Here, for the convenience of calculation, we chose the non-unit mass fields. The topological indexes are consistent with the results obtained in the unit mass fields. These values in III and V surfaces figure out a path, which cross the zero point, i.e., 
\begin{equation}
\begin{split}
&C^{(1)}_0=C_\text{III,V}=\frac{1}{2}\left[\text{sgn}(\mathbf{M}^{(2,1)}_\text{III})-\text{sgn}(\mathbf{M}^{(2,1)}_\text{V})\right]=1,\\
&C^{(2)}_0=C_\text{III,V}=\frac{1}{2}\left[\text{sgn}(\mathbf{M}^{(2,2)}_\text{III})-\text{sgn}(\mathbf{M}^{(2,2)}_\text{V})\right]=1,
\end{split}
\end{equation}
as shown in Fig.~\ref{Fig:s1}(b) and \ref{Fig:s1}(c). Hence we
have $\mathbf{C}_\text{III,V}=\left(C^{(1)}_0,C^{(2)}_0\right)=(1,1)$, implying that there are topological zero-energy states in this edge. For example of the edge intersected V and VI surfaces, we obtain the non-unit mass field as 
\begin{equation}
\begin{split}
\mathbf{M}^{(2,1)}_\text{V}=-\frac{B_2}{\sqrt{B_2^2+m_0^2}},~\mathbf{M}^{(2,1)}_\text{VI}=-\frac{B_1}{\sqrt{B_1^2+m_0^2}},~\mathbf{M}^{(2,2)}_\text{V}=-\frac{m_0}{\sqrt{B_2^2+m_0^2}},~\mathbf{M}^{(2,2)}_\text{VI}=\frac{m_0}{\sqrt{B_1^2+m_0^2}}.
\end{split}
\end{equation}
These values in V and VI surfaces figure out a path, which can not cross the zero point, i.e., 
\begin{equation}
\begin{split}
&C^{(1)}_0=C_\text{V,VI}=\frac{1}{2}\left[\text{sgn}(\mathbf{M}^{(2,1)}_\text{V})-\text{sgn}(\mathbf{M}^{(2,1)}_\text{VI})\right]=0,\\
&C^{(2)}_0=C_\text{V,VI}=\frac{1}{2}\left[\text{sgn}(\mathbf{M}^{(2,2)}_\text{V})-\text{sgn}(\mathbf{M}^{(2,2)}_\text{VI})\right]=-1,
\end{split}
\end{equation}
as shown in Fig.~\ref{Fig:s1}(b) and \ref{Fig:s1}(c). Hence we
have $\mathbf{C}_\text{V,VI}=\left(C^{(1)}_0,C^{(2)}_0\right)=(0,-1)$, implying that there is no zero-energy state in this edge. Similarly, we can identify $C^{(1),(2)}_0$ for the remaining edges, as shown in Fig.~\ref{Fig:s1}(b). Finally, the $3$D inversion-symmetric HyOTI are determined, where the helical states are located at two edges along the $z$ direction and the corner states are located at one corner of the other two edges; see Fig.~\ref{Fig:s1}(b). These results completely match with the OBC energy spectrum of $\mathcal{H}(\mathbf{k})$ in the main text.}

\begin{figure}[!t]
\centering
\includegraphics[width=0.98\columnwidth]{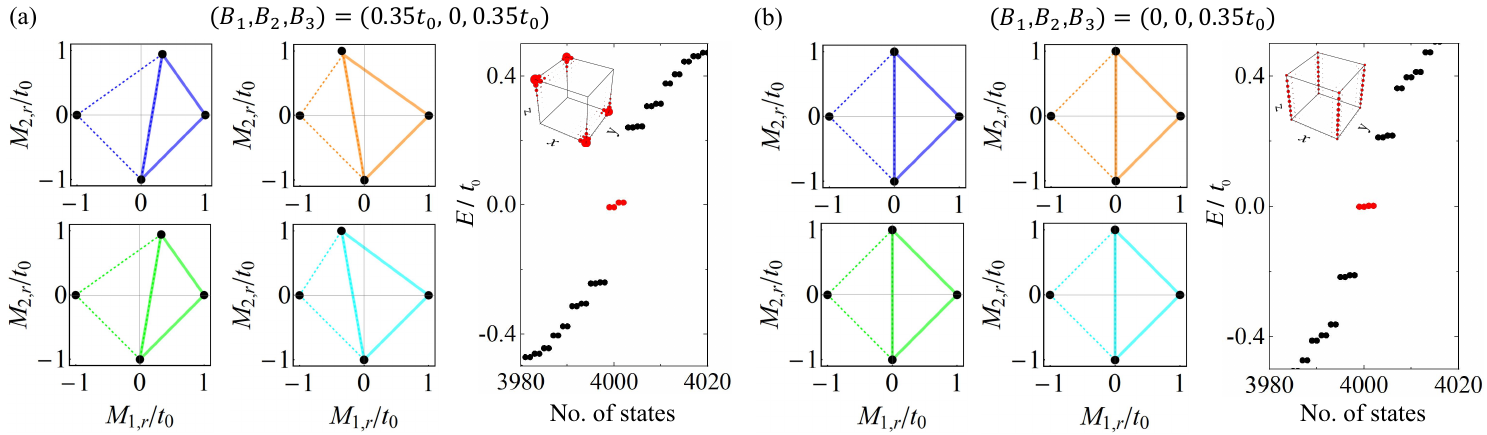}
\caption{Distribution of the unit mass fields enclosing 8 corners and 12 edges of (a) for $(B_1,B_2,B_3)=(0.35t_0,0,0.35t_0)$ and (b) for $(B_1,B_2,B_3)=(0,0,0.35t_0)$. The corresponding OBC energy spectrum gives four zero-energy states located in 4 corners for (a) and 4 hinges for (b). Here we take $m_0=5t_0$ and the lattice size is $N_x=N_y= N_z=10$. 
}
\label{Fig:s2}
\end{figure}

Besides, we emphasize that this surface theory is also applicable to determine higher-order topological states. In the model \eqref{s_model}, the higher-order topological phases can be induced by adjusting the parameters. When taking $(B_1,B_2,B_3)=(0.35t_0,0,0.35t_0)$, we calculate the unit mass fields for all corners and edges, as shown in Fig.~\ref{Fig:s2}(a). It reveals a third-order topological phase with corner states. These corner states are located at the left (right) upper and left (right) lower of $(001)$ [$(00\bar{1})$] surface. The OBC energy spectrum of $\mathcal{H}(\mathbf{k})$ further confirms these results. Moreover, we take $(B_1,B_2,B_3)=(0,0,0.35t_0)$ to calculate the unit mass fields for all corners and edges, as shown in Fig.~\ref{Fig:s2}(b). It reveals a second-order topological phase with helical states, which are located at four edges along $z$ direction. Similarly, the OBC energy spectrum of $\mathcal{H}(\mathbf{k})$ confirms these results. These results demonstrate that the surface theory is a powerful tool for determining the orders of the hybrid-order topology. 

\subsection{IV. Circuit-based realization}

We next show the expression of circuit Laplacian $\mathcal{J}(\omega)$ in detail. For the cubic circuit of Fig. 3 in the main text, the frequency-dependent Laplacian $\mathcal J(\omega)$ takes the form
\begin{equation}
\mathcal{J}(\omega)=\begin{bmatrix}
j_{11} & j_{12} & j_{13} & j_{14} & j_{15} & 0 & 0 & 0 \\
j_{21} & j_{22} & j_{23} & j_{24} & 0 & j_{26} &  0 & 0 \\
j_{31} & j_{32} & j_{33} & j_{34} & 0 & 0  & j_{37} & 0 \\
j_{41} & j_{42} & j_{43} & j_{44} & 0 & 0 & 0 & j_{48} \\
-j_{15} & 0 & 0 & 0 & -j_{11} & -j_{12} & -j_{13} & -j_{14} \\
0 & -j_{26} & 0 & 0 & -j_{21} & -j_{22} & -j_{23} & -j_{24} \\
0 & 0 & -j_{37} & 0 & -j_{31} & -j_{32} & -j_{33} & -j_{34} \\
0 & 0 & 0 & -j_{48} & -j_{41} & -j_{42} & -j_{43} & -j_{44}
\end{bmatrix}.
\end{equation}
The matrix elements are given by
$j_{11}=-6\omega C_4+\omega C_4\sum_i\cos k_i-\omega C_B+\mathtt{i}G_B-\mathtt{i}G_{ii}-1/(\omega L_g)+\omega C_{ii}$,
$j_{12}=\omega C_B-\mathtt{i}G_B$,
$j_{13}=2G_3\sin k_3$,
$j_{14}=2G_1\sin k_1-2\mathtt{i}\omega C_2\sin k_2$,
$j_{15}=-2\mathtt{i}G_5(\cos k_1-\cos k_2)$.
By carefully selecting the values of $C_{ii}$ and $G_{ii}$, we obtain $j_{11}=\omega C_g-1/(\omega L_g)+\omega C_4\sum_i\cos k_i$, which can provide desired on-site potentials at given frequency $\omega_0$. Using the same design principles, the complete Hamiltonian for the inversion-symmetric HyOTI can be implemented. It is noted that the sign of the hoppings can be controlled by the directions of INIC and the signs of the capacitors [a negative capacitor is equivalent to an inductor because $\mathtt{i}\omega C=-\mathtt{i}/(\omega L)$ at $\omega=1/\sqrt{LC}$].

\begin{figure}[!t]
\centering
\includegraphics[width=0.98\columnwidth]{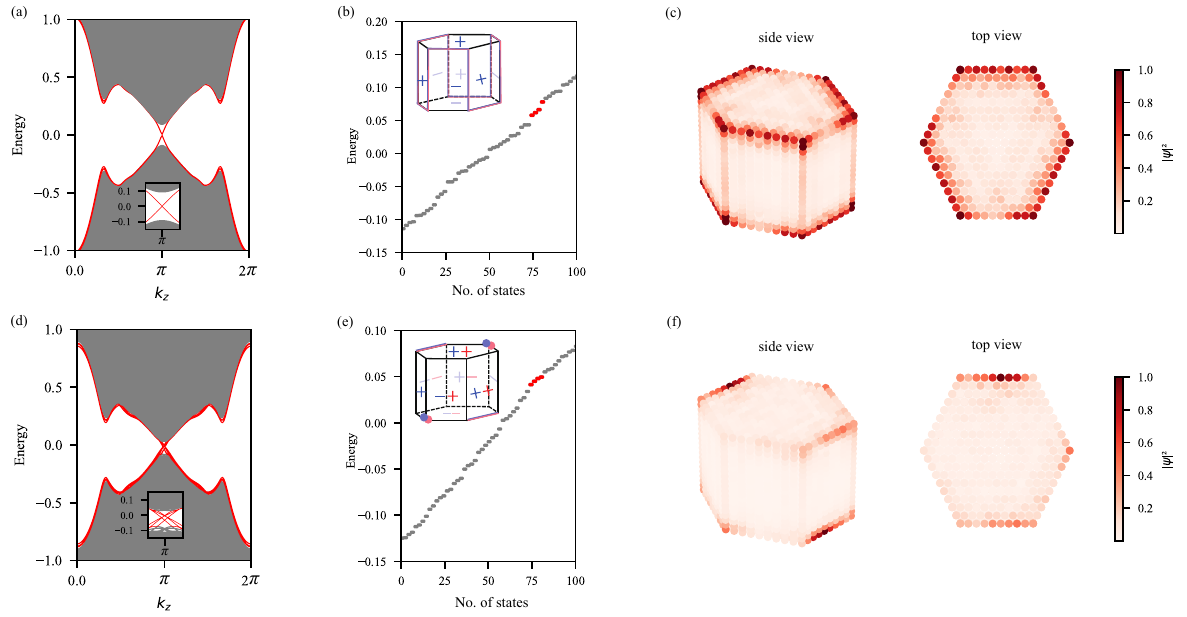}
\caption{Numerical results of the tight-binding Hamiltonian $\mathcal{H}_\text{TB}(\mathbf{k})$ on a hexagonal geometry. The $k_z$-resolved energy spectrum (a) without and (d) with additional mass terms, where the system size is $52\times 52$. The insets give the details of energy bands near $k_z=\pi$, where we take the system size as $72 \times 72$. The 100 eigenstates near the Fermi level are plotted by taking the completely OBCs, where two insets illustrate the signs of effective surface masses, respectively. Here we take the system size as $20\times 20\times 20$. The corresponding real-space contributions of the states with red color are shown in (c) and (f), respectively. Both side and top views show a second-order topological phase for (c) and a unconventional hybrid-order (coexistence of helical and corner states) topological phase for (f).}
\label{Fig:s3}
\end{figure}

\subsection{V. Tight-binding model in a 3D hexagonal lattice}

We next provide a tight-binding model based on a $3$D hexagonal lattice to realize the uncoventional HyOTI with both helical and corner states. Firstly, we consider a tight-binding model $\mathcal{H}_\text{Bi}(\mathbf{k})$, which is topologically equivalent to a realistic model of bismuth \cite{PhysRevB.52.1566,NJP123015}. It has three-fold rotation symmetry $C^{z}_{3}$ and inversion symmetry $\mathcal{I}$ and is defined on the $3$D hexagonal lattice spanned by the lattice vectors $\textit{\textbf{a}}_1=(1,0,0),\textit{\textbf{a}}_2=(-\frac{1}{2},\frac{\sqrt{3}}{2},0)$ and $\textit{\textbf{a}}_3=(0,0,1)$~\cite{10.1038/s41567-018-0224-7}. The corresponding Bloch Hamiltonian has 8 bands and is given by
\begin{equation}
    \mathcal{H}_\text{Bi}(\mathbf{k})=
    \begin{bmatrix}
    \mathcal{H}_\text{I}(\mathbf{k}) & \delta \mathcal{M}(\mathbf{k})\\
    \delta \mathcal{M}(\mathbf{k})^{\dagger} & \mathcal{H}_\text{II}(\mathbf{k}) \\
    \end{bmatrix}.
\end{equation}
This Hamiltonian $\mathcal{H}_\text{Bi}(\mathbf{k})$ consists of two 3D topological insulators given by $\mathcal{H}_\text{I}(\mathbf{k})$ and $\mathcal{H}_\text{II}(\mathbf{k})$, which belong to the different $C_3^z$ subspaces and are couple together via the mass matrix $\mathcal{M}(\mathbf{k})$ with the coupling strength $\delta$. Moreover, we have
\begin{equation}
    \begin{split}
\mathcal{H}_\text{I}(\mathbf{k})=&\Gamma_1 \left \{ m_\text{I}(1+\mathrm{cos}\mathbf{k}\cdot \textit{\textbf{a}}_3)-t_\text{I}[\mathrm{cos}\mathbf{k}\cdot \textit{\textbf{a}}_1+\mathrm{cos}\mathbf{k}\cdot \textit{\textbf{a}}_2+\mathrm{cos}\mathbf{k}\cdot(\textit{\textbf{a}}_1+\textit{\textbf{a}}_2)] \right \} \\
    &+ \lambda_\text{I}[\Gamma_2 \mathrm{sin}\mathbf{k}\cdot \textit{\textbf{a}}_1+\Gamma_{2,1}^\text{I,I}\mathrm{sin}\mathbf{k}\cdot \textit{\textbf{a}}_2
    -\Gamma_{2,2}^\text{I,I} \mathrm{sin}\mathbf{k}\cdot(\textit{\textbf{a}}_1+\textit{\textbf{a}}_2)
    +\Gamma_3\mathrm{sin}\mathbf{k}\cdot \textit{\textbf{a}}_3],
    \end{split}
\end{equation}  
\begin{equation}
\begin{split}
       \mathcal{H}_\text{II}(\mathbf{k})=&\Gamma_1\left \{m_\text{II}(1+ \mathrm{cos}\mathbf{k}\cdot \textit{\textbf{a}}_3)-t_\text{II}[\mathrm{cos}\mathbf{k}\cdot \textit{\textbf{a}}_1+\mathrm{cos}\mathbf{k}\cdot \textit{\textbf{a}}_2+\mathrm{cos}\mathbf{k}\cdot(\textit{\textbf{a}}_1+\textit{\textbf{a}}_2)] \right \}\\
       &+\lambda_\text{II}[\Gamma_2 \mathrm{sin}\mathbf{k}\cdot \textit{\textbf{a}}_1
       +\Gamma_{2,1}^\text{II,II}\mathrm{sin}\mathbf{k}\cdot \textit{\textbf{a}}_2
       -\Gamma_{2,2}^\text{II,II} \mathrm{sin}\mathbf{k}\cdot(\textit{\textbf{a}}_1+\textit{\textbf{a}}_2)
        +\Gamma_3\mathrm{sin}\mathbf{k}\cdot \textit{\textbf{a}}_3]\\
        &+\Gamma_4 \gamma_\text{II}[\mathrm{sin}\mathbf{k}\cdot(\textit{\textbf{a}}_1+2\textit{\textbf{a}}_2)
        +\mathrm{sin} \mathbf{k}\cdot(\textit{\textbf{a}}_1-\textit{\textbf{a}}_2)
        -\mathrm{sin} \mathbf{k}\cdot(2\textit{\textbf{a}}_1+\textit{\textbf{a}}_2)],
    \end{split}
\end{equation}
\begin{equation}
    \begin{split}
       \mathcal{M}(\mathbf{k})=&\Gamma_2[\mathrm{sin}\mathbf{k}\cdot \textit{\textbf{a}}_1
       +\mathrm{sin} \mathbf{k}\cdot(2\textit{\textbf{a}}_1+\textit{\textbf{a}}_2)] 
       +\Gamma_{2,1}^\text{I,II}[\mathrm{sin}\mathbf{k}\cdot \textit{\textbf{a}}_2
       +\mathrm{sin} \mathbf{k}\cdot(\textit{\textbf{a}}_2-\textit{\textbf{a}}_1)]\\
       &-\Gamma_{2,2}^\text{I,II}[\mathrm{sin}\mathbf{k}\cdot(\textit{\textbf{a}}_1+\textit{\textbf{a}}_2)
       +\mathrm{sin} \mathbf{k}\cdot(\textit{\textbf{a}}_1+2\textit{\textbf{a}}_2)]
       -\mathtt{i}\Gamma_5[\mathrm{cos}\mathbf{k}\cdot \textit{\textbf{a}}_1
       +\mathrm{cos} \mathbf{k}\cdot(2\textit{\textbf{a}}_1+\textit{\textbf{a}}_2)]\\
       &-\mathtt{i}\Gamma_{5,1}^\text{I,II}[\mathrm{cos}\mathbf{k}\cdot \textit{\textbf{a}}_2
       +\mathrm{cos} \mathbf{k}\cdot(\textit{\textbf{a}}_2-\textit{\textbf{a}}_1)]
       -\mathtt{i}\Gamma_{5,2}^\text{I,II}[\mathrm{cos} \mathbf{k}\cdot(\textit{\textbf{a}}_1+\textit{\textbf{a}}_2)+\mathrm{cos} \mathbf{k}\cdot(\textit{\textbf{a}}_1+2\textit{\textbf{a}}_2)],
    \end{split}
\end{equation}
where $\Gamma_1=\tau_3\sigma_0, \Gamma_2=\tau_1 \sigma_1,\Gamma_3=\tau_2\sigma_0,\Gamma_4=\tau_1\sigma_2,\Gamma_5=\tau_3\sigma_1$, and $\Gamma_{\mu,\nu}^{i,j}=(C_{3,i}^z)^\nu\Gamma_\mu (C_{3,j}^z)^{-\nu}$. Here we have $\mu \in\left \{ 1,\cdots,5 \right \}$, $i,j \in\left \{\text{I,II}\right \}$, $\nu \in \left \{ 1,2 \right \}$ , $C_{3,\text{I}}^z=\tau_0\text{e}^{\mathtt{i}\frac{\pi}{3}\sigma_3}$, and $C_{3,\text{II}}^z=-\tau_0\sigma_0 $, so that the threefold rotation symmetry is given by $C_3^z=C_{3,\text{I}}^z\oplus C_{3,\text{II}}^z$. It is noted that the Pauli matrices $\sigma_i$ and $\tau_i$ denote the freedom degree of spin and orbit, respectively. By taking the parameter $m_\text{I}=m_\text{II}=2.0$, $t_\text{I}=t_\text{II}=1.0$, $\lambda_\text{I}=0.3$, $\lambda_\text{II}=\gamma_\text{II}=1.0$, and $\delta=0.3$, we obtain the $k_z$-resolved energy spectrum in Fig.~\ref{Fig:s3}(a). There are six Kramers pairs in the gapless helical states [see the inset of Fig.~\ref{Fig:s3}(a)], implying that there are helical states in the $6$ edges of the $3$D hexagonal geometry. Further, we take the OBCs along all directions and obtain its energy spectrum, of which 100 states near the Fermi level are plotted in Fig.~\ref{Fig:s3}(b). The real-space contributions of 8 states marked as red color are shown in Fig.~\ref{Fig:s3}(c), confirming the second-order topological phases with helical modes. Such topological states can be understood through the picture of mass domain wall induced by the effective surface masses; see the inset of Fig.~\ref{Fig:s3}(b). 

Next, we introduce two additional mass terms $m_{x,z}$ into the tight-binding model of bismuth. The total Hamiltonian is given by 
\begin{equation}
\mathcal{H}_\text{TB}(\mathbf{k})=\mathcal{H}_\text{Bi}(\mathbf{k})+\mathcal{H}_\text{m}(\mathbf{k}),~~ \mathcal{H}_\text{m}(\mathbf{k})=
    \begin{pmatrix}
    m_z\tau_0 \sigma_0 & m_x\tau_0 \sigma_0\\
    m_x\tau_0 \sigma_0& -m_z\tau_0 \sigma_0\\
    \end{pmatrix}.
\end{equation}
As a result, these helical states along $z$ direction open a energy gap through the nonzero $m_{x,z}$. By taking the parameters $m_z=-0.05$ and $m_x=-0.15$, it is seen that the degeneracy of six Kramers pairs are broken in Fig.~\ref{Fig:s3}(d). Meanwhile, a very small band gap emerge; see the inset of Fig.~\ref{Fig:s3}(d). Since $m_{x,z}$ provide the additional surface masses, the effective mass fields are changed into the forms of the inset of Fig.~\ref{Fig:s3}(e). Applying the proposed surface theory, we immediately determine that the system has the coexistence of 0D corner states and 1D edge states. The OBC energy spectrum in Figs.~\ref{Fig:s3}(e) and \ref{Fig:s3}(f) confirm these results.  

\end{document}